\documentclass[12pt]{article}
\usepackage[T1]{fontenc}
\usepackage[latin1]{inputenc}

\usepackage{epsfig}

\newcommand{\lsim}{\,{\buildrel < \over {_\sim}}\,}
\newcommand{\gsim}{\,{\buildrel > \over {_\sim}}\,}

\begin{document}

\begin{titlepage}
\begin{flushright}
HIP-2003-47/TH\\
LBNL-53839\\
hep-ph/0310111\\
9 October 2003
\end{flushright}
\vfill
\begin{centering}

{\bf ENHANCEMENT OF CHARM QUARK PRODUCTION DUE TO NONLINEAR CORRECTIONS
TO THE DGLAP EQUATIONS}\\

\vspace{0.5cm}
K.J. Eskola$^{\rm a,b}$\footnote{kari.eskola@phys.jyu.fi},
V.J. Kolhinen$^{\rm a}$\footnote{vesa.kolhinen@phys.jyu.fi} and
R. Vogt$^{\rm a,c}$\footnote{vogt@lbl.gov}

\vspace{1cm}
{\em $^{\rm a}$Department of Physics, University of Jyv\"askyl\"a,\\
P.O. Box 35, FIN-40014 University of Jyv\"askyl\"a}
\vspace{0.3cm}

{\em $^{\rm b}$Helsinki Institute of Physics,\\
P.O. Box 64, FIN-00014 University of Helsinki, Finland}
\vspace{0.3cm}

{\em $^{\rm c}$Lawrence Berkeley National Laboratory, Berkeley, CA 94720, 
USA,\\  and \\ Physics Department, 
University of California, Davis, CA 95616,
USA}

\vspace{0.3cm}

\vspace{1cm} 
{\bf Abstract} \\ 
\end{centering}

We have studied how parton distributions based on the inclusion of
nonlinear scale evolution and constraints from HERA data affect charm
production in $pp$ collisions at center-of-mass energies of 5.5, 8.8
and 14 TeV.  We find that, while the resulting enhancement can be
substantial, it is very sensitive to the charm quark mass and the
scale entering the parton densities and the strong coupling constant.

\vspace{0.3cm}\noindent

\vfill
\end{titlepage}

\section{Introduction}

Global fits of the parton distribution functions (PDFs) such as those
by CTEQ \cite{CTEQ6,cteq61} and MRST
\cite{MRST2001,MRSTNNLO,Martin:2003sk}, based on the
Dokshitzer-Gribov-Lipatov-Altarelli-Parisi (DGLAP) \cite{DGLAP} scale
evolution, successfully describe the proton structure function,
$F_2(x,Q^2)$, deep inelastic scattering (DIS) data in the ``high
$(Q^2,x)$'' region, $Q^2 \gsim 10$ GeV$^2$ and $x \gsim 0.005$.
However, it is not possible to maintain the excellent high $(Q^2,x)$
DIS fit while simultaneously fitting the ``low $(Q^2,x)$'' region,
$1.5 \lsim Q^2 \lsim 10$~GeV$^2$ and $x\lsim 0.005$
\cite{Martin:2003sk}.  In addition, the next-to-leading order (NLO)
gluon distribution becomes negative for sufficiently small $x$ at the
few GeV$^2$ scales.

Nonlinear corrections to the PDF evolution based on gluon
recombination were first derived by Gribov, Levin and Ryskin
\cite{GLR} as well as Mueller and Qiu \cite{MQ}. Recent work
\cite{EHKQS} showed that adding these GLRMQ terms to the DGLAP
equations can improve the overall leading order (LO) fits to the HERA
DIS data \cite{H1}.  The rapid $Q^2$ evolution in the low $(Q^2,x)$
region from DGLAP alone is slowed by the GLRMQ recombinations.  At
$Q^2$ scales far above the initial scale $Q_0^2$, the $Q^2$ evolution
of the PDFs is again described by the DGLAP equations since the GLRMQ
terms become negligible, see Fig.~\ref{gluons}.

While the quark distributions are directly constrained by the HERA
$F_2(x,Q^2)$ data, the gluon distribution is constrained by the $F_2$
slope, $\partial F_2/\partial \ln Q^2$.  The fairly modest measured
slopes force the LO DGLAP gluon distributions to be nearly independent
of $x$ for scales of a few GeV$^2$ \cite{CTEQ6,MRST2001} and also
force the NLO gluons to be negative.  When the nonlinear terms are
included, the slowing of the $Q^2$ evolution leads to an enhancement
of the small-$x$ gluon distributions at $Q^2\lsim10$~GeV$^2$
relative to the LO DGLAP gluon distributions, subject to the same
constraints from HERA \cite{EHKQS}.  The effect of nonlinear evolution
on the NLO distributions is not yet fully explored by global fits
which include the low $(Q^2,x)$ region but the results of
Ref.~\cite{Martin:2003sk} suggest that the enhancement is smaller than
at LO.  This effect alone thus seems unlikely to produce positive NLO
small $x$ gluon distributions at the few GeV$^2$ scales.

In spite of the problems described above, the quality of the global
DGLAP fits to the HERA data \cite{CTEQ6,MRST2001} is good.  The
$\chi^2$ per degree of freedom is close to one even when the low
$(Q^2,x)$ region is included.  Therefore, $F_2$ measurements at HERA
alone may not clearly differentiate DGLAP from nonlinear
evolution and more direct probes of the gluon distribution are
needed.  In this paper, we study whether the parton distribution
functions generated with the LO DGLAP+GLRMQ evolution in
Ref.~\cite{EHKQS} could give rise to any significant enhancement
relative to the DGLAP-evolved PDFs in charm quark hadroproduction.  Charm
production is the best candidate process since the charm quark mass is
relatively low, $1.2 \lsim m_c \lsim 1.8$ GeV, and its production is
dominated by gluons.  These two characteristics should lead to the
most favorable conditions for a possible effect.  The $Q^2$ scale at
which the total cross sections are calculated is proportional to
$m_c^2$ and $4m_c^2$.  Thus the results are sensitive to
$Q^2$. Unfortunately, due to the small charm mass, the scale
dependence is still significant at NLO \cite{RVkfac}.

We focus on $pp$ collisions since these nonlinear distributions are
not yet available for nuclei.  Our calculations are at leading order
only since the nonlinear parton distribution functions, referred
to as EHKQS hereafter, are only evolved to LO in Ref.~\cite{EHKQS}.
We calculate the possible effect as a function of rapidity and
transverse momentum of the charm quark and the $c \overline c$ pair
invariant mass.  We study collisions at center-of-mass energies
$\sqrt{S} = 5.5$, 8.8 and 14 TeV at the Large Hadron Collider (LHC).
These energies correspond to the planned per nucleon energies of
Pb+Pb, $p$Pb and $pp$ collisions, effectively spanning the LHC energy
regime.

\section{Formalism and inputs}

\subsection{Cross sections and parton distribution functions}

Inclusive differential charm cross sections at high energies 
are, to first approximation, computable assuming factorization. 
The cross section may be expressed as
\begin{eqnarray}
 d\sigma_{pp\rightarrow c \overline c X}(Q^2,\sqrt S) =
   \sum_{i,j,k=q,\overline q,g} 
   f_i(x_1,Q^2)\otimes f_j(x_2,Q^2)
   \otimes d\hat \sigma_{ij\rightarrow c \overline c k}(Q^2,x_1,x_2)
\label{sigcc}
\end{eqnarray}
where $\hat \sigma_{ij\rightarrow c \overline c k}(Q^2,x_1,x_2)$ are
the perturbatively calculable partonic cross sections for charm
production at scales $Q^2\gg\Lambda^2_{\rm QCD}$, $x_1$ and $x_2$ are
the momentum fractions of the partons involved in the hard scattering
and $f_i(x,Q^2)$ are the free proton parton densities.  At LO, only
the $gg$ and $q \overline q$ channels are available and $k = 0$, {\it
i.e.} no other partons are produced with the $c \overline c$.  The LO
matrix elements and partonic cross sections for charm production can
be found in Ref.~\cite{COMBRIDGE}.  In the following, we consider the
triple differential distributions, $d^3\sigma/(dp_Tdydy_2)$, where $y$
and $y_2$ are the rapidities of the quark and antiquark and $p_T$ is
the quark transverse momentum.  We also study the inclusive
distributions $d\sigma/dy$, $d\sigma/dp_T$ and $d\sigma/dM$ where
$M^2= 2m_T^2 (1 + \cosh(y - y_2))$ is the square of the pair invariant
mass and $m_T^2 = p_T^2 + m_c^2$ is the transverse mass of the quark.

The new inputs in this straightforward calculation are the
nonlinearly-evolved proton PDFs.  The EHKQS sets\footnote{available at
www.urhic.phys.jyu.fi} in Ref.~\cite{EHKQS}
employ the CTEQ5L \cite{CTEQ5} and CTEQ6L \cite{CTEQ6} PDFs as a
baseline and require a good fit to the HERA $F_2^p(x,Q^2)$ data over
the full $(Q^2,x)$ range \cite{EHKQS}.  Three EHKQS sets were obtained
when the GLRMQ terms were included in the analysis.  All three EHKQS
sets employ the initial scale $Q_0^2 = 1.4$ GeV$^2$ and $\Lambda_{\rm
QCD} = 0.192$ GeV for four flavors but differ in the treatment of the
charm mass threshold in the evolution.  In set 1, a nonzero charm
distribution was allowed at $Q_0^2$.  Sets 2a and 2b assumed that
there was no charm quark distribution below $Q_0^2$, turning on when
$Q^2 = m_c^2 \geq Q_0^2$.  Set 2a assumed $m_c = 1.3$ GeV while set 2b
took $m_c = Q_0 = \sqrt{1.4}$ GeV.  The input gluon distribution is
the same in all cases so that the small differences at high $Q^2$
arise from the treatment of the charm quark evolution.  The choice of
set 1, set 2a or set 2b therefore makes very little difference in the
overall effect of the nonlinear terms on charm production.  Thus we
only use EHKQS set 1.

We work at leading order since the EHKQS sets are only evolved to LO
using a one-loop evaluation of the strong coupling constant
$\alpha_s$.  Thus these LO distributions should generally not be mixed
with NLO matrix elements and the two-loop $\alpha_s$.  The charm quark
$p_T$ distribution is broadened at NLO relative to the LO calculation
\cite{RVkfac}.  Therefore we study the ratios of calculations with EHKQS
relative to those with a standard LO PDF set evolved using the DGLAP
equations alone.  We quantify the effect with respect to the
CTEQ61L parameterization, the most recent LO fit to the PDFs that also
uses a one-loop $\alpha_s$ \cite{cteq61}.  The minimum scale of
CTEQ61L is $Q_0^2 = 1.69$ GeV$^2$.  This LO fit obtained a slightly
higher value of $\Lambda_{\rm QCD}$ for four flavors, $0.215$ MeV.  In
our CTEQ61L calculations, for consistency with the PDF set, we use this
value in $\alpha_s$.

\subsection{Comparison of EHKQS and CTEQ61L gluon distributions}

Before presenting our results, it is instructive to discuss the
differences between the EHKQS and CTEQ61L gluon distributions in more
detail.  Since the high $\sqrt{S}$ collisions studied here probe the
very low $x$ region, some remarks on the region of applicability of
the PDFs are in order.  Below the minimum $x$ and $Q^2$ values assumed
in the fits, the PDFs are essentially unconstrained.  At the scales
studied here, described in the following section, $Q^2$ always remains
above the minimum scale of the PDF sets.  However, the region below
the minimum $x$, $x_{\rm min}$, is reached at large rapidities and
intermediate scales at the LHC.  Thus the behavior of the PDFs below
$x_{\rm min}$ is an uncertainty for all PDFs.  This situation can
only be improved by more extensive small $x$ constraints on the PDFs.

The minimum $x$ of the EHKQS sets is $x_{\rm min}^{\rm EHKQS} = 10^{-5}$.
For $x<x_{\rm min}^{\rm EHKQS}$ and $Q^2$ of a few GeV$^2$, neglected
power-suppressed terms in the evolution become important and the
DGLAP+GLRMQ results are no longer trustworthy \cite{EHKQS}.  The
CTEQ61L minimum is an order of
magnitude smaller than $x_{\rm min}^{\rm EHKQS}$, 
$x_{\rm min}^{\rm CTEQ61L} = 10^{-6}$.  The very small $x$
regions below $x_{\rm min}$ are not excluded from our calculations.
Instead, we assume that below $x_{\rm min}$, $f_i(x<x_{\rm
min},Q^2)=f_i(x_{\rm min},Q^2)$ for each set.  We note that in the
CTEQ61L table \cite{cteqwww} the distributions are not constant below
$x_{\rm min}^{\rm CTEQ61L}$.

\begin{figure}[htb]
\vspace{-1.0cm}
\centering\includegraphics[width=10cm]{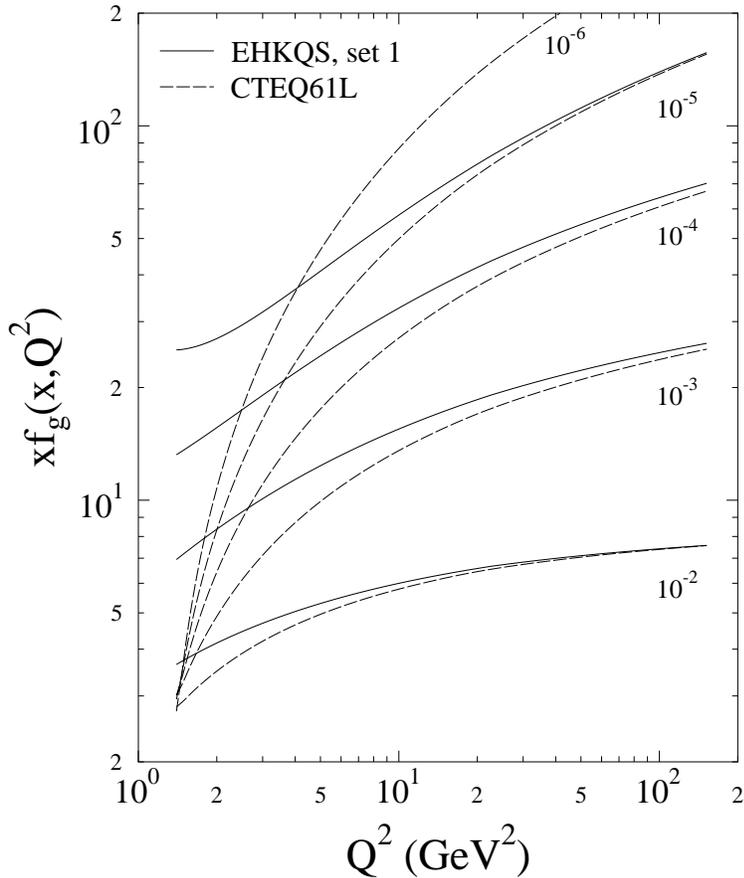}
\vspace{-1.0cm}
\caption[]{\small Comparison of the EHKQS set 1 (solid curves) and CTEQ61L 
(dashed curves) gluon distributions as
a function of $Q^2$ for, from lowest to highest, $x = 10^{-2}$, $10^{-3}$,
10$^{-4}$, $10^{-5}$ and, for CTEQ61L only, $10^{-6}$.
}
\label{gluons}
\end{figure}

It is illustrative to compare the EHKQS set 1 and CTEQ61L gluon
distributions as a function of $Q^2$ for several values of $x$, shown
in Fig.~\ref{gluons}.  Due to the nonlinear evolution, the $Q^2$
dependence of EHKQS set 1 is rather mild compared to CTEQ61L which
approaches a constant at $Q_0^2$ and $x \rightarrow 0$.  Note that for
all $x > 10^{-5}$, the EHKQS distributions are always above CTEQ61L
although, at higher scales, the distributions lie very close together.
In the unconstrained region where $x < 10^{-5}$, the situation clearly
depends on how the extrapolation towards $x\rightarrow0$ is done.  The
CTEQ61L parameterization continues to $x_{\rm min}^{\rm CTEQ61L}$.  In
this very low $x$ region, at $Q^2 = 4$ GeV$^2$, the CTEQ61L gluon
distribution at $x_{\rm min}^{\rm CTEQ61L}$ crosses the EHKQS
distribution, fixed at the value of $x_{\rm min}^{\rm EHKQS}$, and
continues to rise.  Therefore, the behavior of the relative kinematics
distributions we compute can be very sensitive to the treatment of the
unconstrained $x$ region.  Since the two distributions are also quite
sensitive to the scale, the ratios we compute will also be strongly
scale dependent.

\subsection{Scale choice}

In our calculations, we use values of the charm quark mass and scale
that have been fit to the total cross section data using NLO
calculations.  The total cross section data cannot be fit by adjusting
$m_c$ and $Q^2$ with a full LO calculation, employing LO PDFs and the
one-loop $\alpha_s$, because the resulting $m_c$ would be too small
for perturbative applications.  See Ref.~\cite{RVkfac} for more
discussion.  The best agreement with the total cross section data is
obtained with $m_c = 1.2$ GeV and $Q^2 = 4m_c^2$ for standard
DGLAP-evolved NLO PDFs such as CTEQ6M \cite{cteq61} and MRST
\cite{mrst}.  Nearly equivalent agreement may be obtained with $m_c =
1.3$ GeV and $Q^2 = m_c^2$ \cite{HPC,rvww02}.  Thus our main results
are based on these inputs.  Alternative fits to the fixed-target total
cross sections can be achieved with larger values of $m_c$ by
separating the factorization scale, $Q_F$, from the renormalization
scale, $Q_R$, e.g. $Q_F^2 = 4m_c^2$ and $Q_R^2 = m_c^2$ or $m_c^2/4$,
and allowing the fast running of $\alpha_s$ in this $Q_R^2$ region to
increase the cross sections.  Note that if $Q^2 = Q_F^2 \leq Q_0^2$,
the PDFs are unconstrained in $Q^2$.  We keep $Q_F^2 = Q_R^2$, as in
all typical PDF fits such as Refs.~\cite{cteq61,mrst}, limiting
ourselves to relatively small values of $m_c$ to obtain agreement with
the total cross section data.  The PDFs are thus evaluated above
$Q_0^2$ for the masses and scales we use.

Note that we have discussed scales proportional to $m_c^2$ in the
calculations of the total cross sections.  Such scales are used
because the total NLO partonic cross section can be written
analytically as a function of $4m_c^2/s$ where $s$ is the partonic
center-of-mass energy squared \cite{NDE}.  In this case, the charm
quark mass in the only relevant scale.  However, in inclusive
distributions such as we compute here, the quark $p_T$ also enters
since a scale proportional to $m_T^2$ is needed to control
$p_T$-dependent logarithms.  Therefore, in our calculated ratios of
distributions, we take $Q^2 = m_T^2$ with $m_c = 1.3$ GeV and $4m_T^2$
with $m_c = 1.2$ GeV.

Whether the high and low energy behavior of the charm cross section
can be described simultaneously by the same values of $m_c$ and $Q^2$
is an open question.  Since the slopes of the gluon distributions
differ at low and high $x$, the scale dependence is a strong function
of $\sqrt{S}$.  At high $x$, $xf_g(x,Q^2)$ is larger at low $Q^2$ than
at higher scales.  Thus at fixed target energies, $\sigma(m_c^2) >
\sigma(4m_c^2)$.  At collider energies, such as at the LHC, $x$ is
small and $xf_g(x,Q^2)$ is increasing with $Q^2$ so that
$\sigma(4m_c^2) > \sigma(m_c^2)$ even though $\alpha_s(m_c^2) >
\alpha_s(4m_c^2)$.  We thus extend the parameter space of our
calculations to study the possible effect on higher masses, $m_c=1.8$
GeV with $Q^2 = m_T^2$ and $4m_T^2$.

\section{Results}

We start with results insensitive to the unconstrained region at
$x<x_{\rm min}=10^{-5}$.  In inclusive kinematics with an identified
charm quark and fixed $x_T=2m_T/\sqrt S$, the unconstrained $x$-region
contributes to charm production at high rapidities, in the region
\begin{equation}
y_u\equiv\ln \left(1/x_T-\sqrt{1/x_T^2-1/x_{\rm min}}\,\right) \le |y| \le
\ln \left(1/x_T+\sqrt{1/x_T^2-1/x_{\rm min}}\,\right) \, \, .
\label{ylimits}
\end{equation}
The upper limit, close to the phase space boundary, is not of interest
here. Expanding the lower limit, $y_u$, in powers of $x_T^2/x_{\rm
min} \ll 1$, we arrive at $y_u \approx \ln [m_T/(x_{\rm min} \sqrt S)]
\ge \ln [m_c/(x_{\rm min}\sqrt S)]$.  If $m_c=1.2$~GeV, the small $x$
region contributes to charm production at $|y|\gsim y_u = 2.2$, 2.6
and 3.1 for $\sqrt S = 14$, 8.8 and 5.5 TeV, respectively.  If
$m_c=1.8$~GeV, the unconstrained region is probed when $|y|\gsim y_u =
2.6$, 3.0 and 3.5, respectively.

\begin{figure}[htb]
\centering\includegraphics[width=13cm]{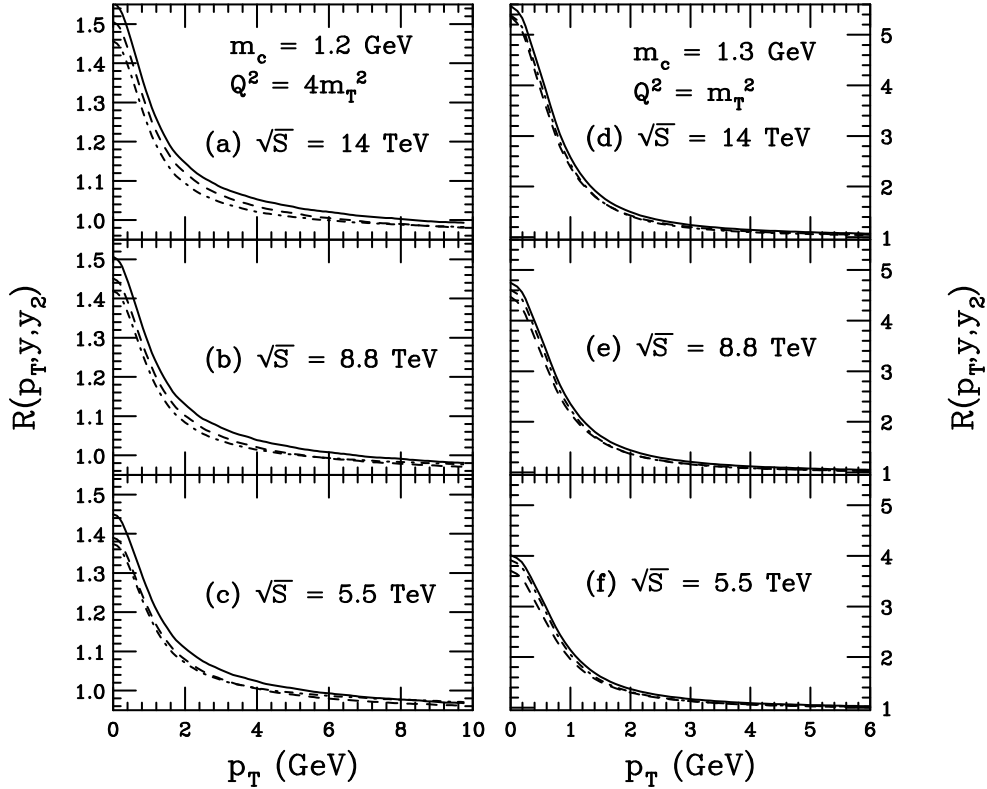}
\caption[]{\small We present $R(p_T,y,y_2)$ for
fixed $y$ and $y_2$ as a function of charm quark $p_T$ at $\sqrt{S}=
14$ TeV, (a) and (d), 8.8 TeV, (b) and (e), and 5.5 TeV, (c) and (f),
in $pp$ collisions.  The results are shown for $y=y_2 = 0$ (solid
curves), $y=2$ $y_2 = 0$ (dashed) and $y=y_2=2$ (dot-dashed).  We show
$m_c = 1.2$ GeV and $Q^2 = 4m_T^2$ on the left-hand side and $m_c = 1.3$
GeV and $Q^2 = m_T^2$ on the right-hand side.  Note the different scales on the
left- and right-hand axes.  }
\label{1213ydep}
\end{figure}

\begin{figure}[htb]
\centering\includegraphics[width=13cm]{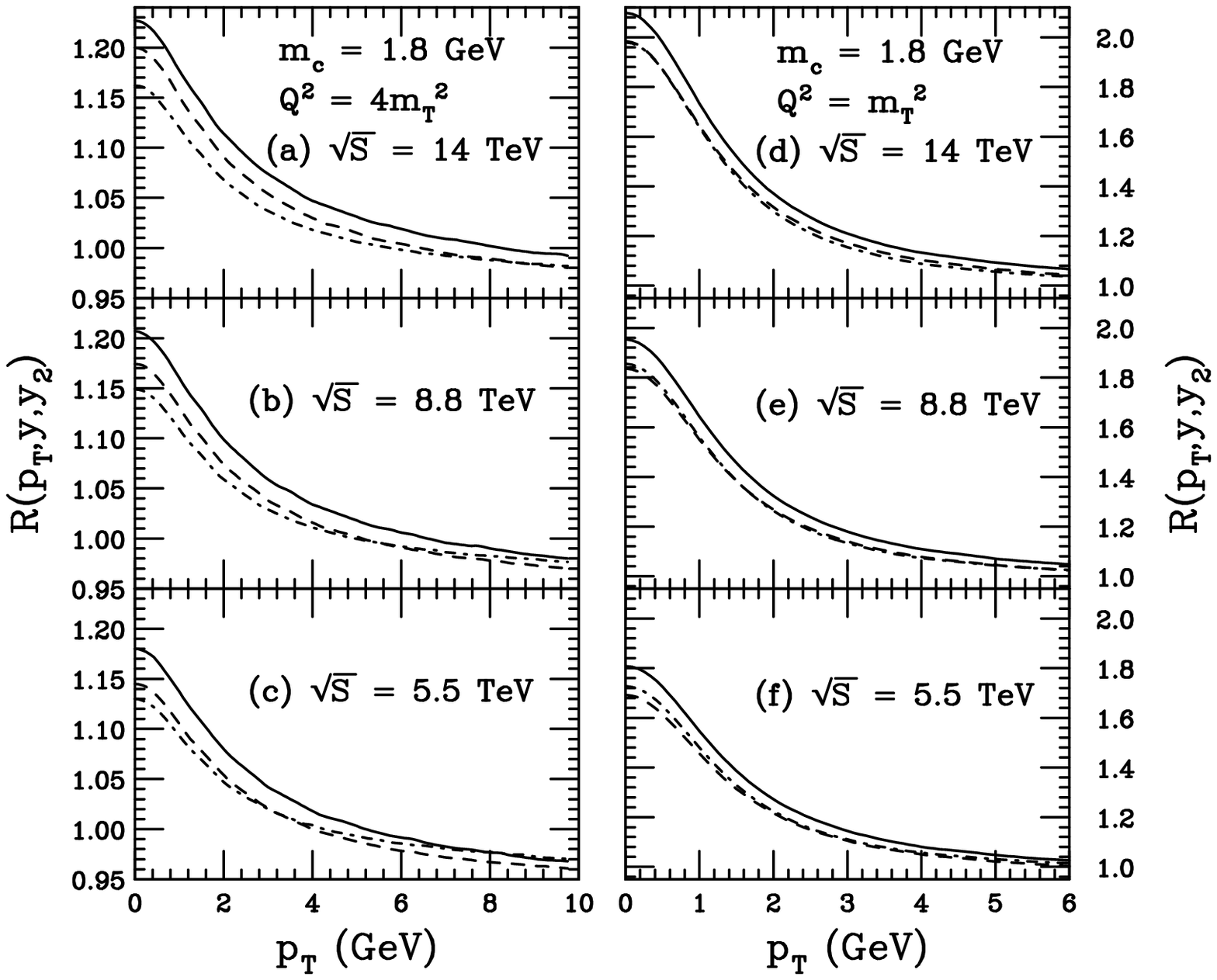}
\caption[]{\small As in Fig.~\protect\ref{1213ydep} but for $m_c = 1.8$ GeV and
$Q^2 = 4m_T^2$ (left-hand side) and for $m_c = 1.8$ GeV and $Q^2 =
m_T^2$ (right-hand side). Note the different scales on the
left- and right-hand axes.
}
\label{18ydep}
\end{figure}

First, we study the ratio of the fully differential cross sections calculated
with the EHKQS densities relative to the CTEQ61L densities at fixed $y$ and
$y_2$,
\begin{equation}
R(p_T,y,y_2) \equiv \frac{d^3\sigma({\scriptstyle\rm EHKQS})/(dp_T dy dy_2)}
{d^3\sigma({\scriptstyle\rm CTEQ61L})/(dp_T dy dy_2)}
\label{rofptyy2}
\end{equation}
In Figs.~\ref{1213ydep} and \ref{18ydep} we present this unintegrated
ratio as a function of $p_T$ for several values of $y$ and $y_2$, all
within the range constrained by data: $y = y_2 = 0$ (solid curves), $y
= 2$, $y_2=0$ (dashed) and $y=y_2 = 2$ (dot-dashed).
Figure~\ref{1213ydep} presents $R(p_T,y,y_2)$ for the values of mass
and scale that agree best with the total cross section data, $m_c =
1.2$ GeV and $Q^2 = 4m_T^2$ on the left-hand side and $m_c = 1.3$ GeV,
$Q^2 = m_T^2$ on the right-hand side.  Figure ~\ref{18ydep} shows the
same unintegrated ratios for our upper limit on $m_c$, 1.8 GeV and
$Q^2 = m_T^2, 4m_T^2$.  The center-of-mass energies shown are
$\sqrt{S} = 14$ (upper), 8.8 (middle) and 5.5 (lower) TeV.  

\begin{figure}[tbh]
\centering\includegraphics[width=14cm]{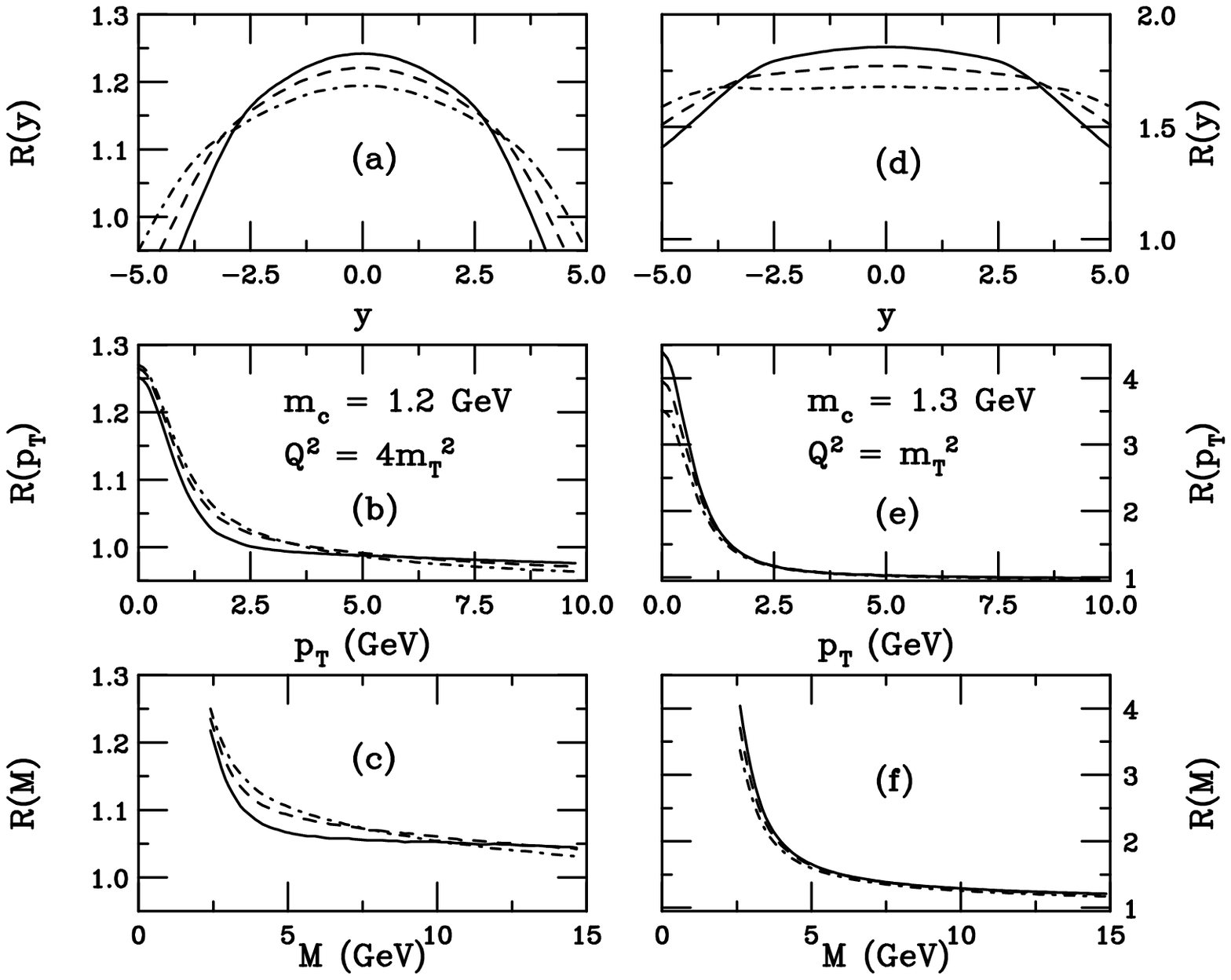}
\caption[]{\small 
We present $R(y)$, (a) and (d), $R(p_T)$, (b) and (e), and $R(M)$,
(c) and (f), in $pp$ collisions at $\sqrt{S} = 14$ (solid), 8.8
(dashed) and 5.5 (dot-dashed) TeV.  The left-hand side shows
$m_c = 1.2$ GeV and $Q^2 = 4m_T^2$, the right-hand side $m_c = 1.3$ GeV 
and $Q^2 = m_T^2$. }
\label{1213edep}
\end{figure}

Figures \ref{1213ydep} and \ref{18ydep}, which demonstrate the
sensitivity of the enhancement to $m_c$ and $Q^2$, can be understood
by inspection of Fig.~\ref{gluons}.  At these high energies, the $gg$
channel dominates so that the partonic cross sections essentially drop
out of the ratios, and
\begin{equation}
R(p_T,y,y_2) \approx 
\frac{x_1 f_g(x_1,Q^2)_{\rm EHKQS}}{x_1 f_g(x_1,Q^2)_{\rm CTEQ61L}} 
\frac{x_2 f_g(x_2,Q^2)_{\rm EHKQS}}{x_2 f_g(x_2,Q^2)_{\rm CTEQ61L}}
\equiv R_g(x_1,Q^2)R_g(x_2,Q^2)\, , 
\label{approxRy}
\end{equation}
where we have denoted the ratio of the EHQKS and CTEQ61L gluons by $R_g$.
The $x$ values are easily calculated at LO for each curve using the
definitions $x_{1,2} = m_T[\exp(\pm y) + \exp(\pm y_2)]/\sqrt{S}$.
Since decreasing $\sqrt{S}$ increases $x$ and consequently decreases
$R_g(x,Q^2)$, the enhancement decreases with energy.  In addition,
$R_g(x,Q^2)$ decreases with increasing $Q^2$, so that increasing $m_c$
and $Q^2$ also reduces the enhancement.  Both $x_1$ and $x_2$ are
small when $y = y_2 = 0$ so that the enhancement is largest at
midrapidity.  Moving away from midrapidity, {\it e.g.} to $y,y_2 > 0$,
increases $x_1$ and decreases $x_2$, correspondingly decreasing
$R_g(x_1,Q^2)$ and increasing $R_g(x_2,Q^2)$.  The rapidity dependence
shown on the right-hand sides of Figs.~\ref{1213ydep} and
\ref{18ydep}, for the lower $Q^2$, is rather weak.  The difference
between the dashed $(y =2, y_2 = 0)$ and dot-dashed $(y=y_2 = 2)$
curves is very small and the enhancement at $y=y_2=2$ lies marginally
above that for $y=2,y_2 =0$ over all $p_T$.  When the scale is small,
the CTEQ61L gluon distribution changes very slowly with $x$ for $x <
0.01$, about 20\% for $10^{-4} \le x \le 10^{-3}$ when $Q^2 = m_c^2$
and $m_c = 1.3$ GeV ($p_T \approx 0$), so that $R(p_T,y,y_2)$ is not a
strong function of $y$ and $y_2$.  However, the results for the larger
scales, shown on the left-hand sides of Figs.~\ref{1213ydep} and
\ref{18ydep}, exhibit the opposite behavior along with a stronger rapidity
dependence.  At larger scales, the slope of the
CTEQ61L gluon distribution with $x$ is considerably stronger, resulting in
a factor of two difference in the gluon distribution
over the range $10^{-4} \le x \le 10^{-3}$ when $Q^2 = 4m_c^2$
and $m_c = 1.2$ GeV ($p_T \approx 0$), introducing a stronger rapidity
dependence of $R(p_T,y,y_2)$ at higher scales.  The nonlinearities die
out at large scales since the EHKQS gluons become similar to the
CTEQ61L gluons so that the enhancement in $R(p_T,y,y_2)$ disappears at
large $p_T$.  The ratio does not become equal to 1 because CTEQ61L and
the EHKQS sets are not identical either at high $Q^2$ or larger
$x$. Note that 
$\alpha_s^2({\scriptstyle{\rm EHKQS}})/\alpha_s^2({\scriptstyle{\rm CTEQ61L}})
\approx 0.9$, allowing $R(p_T,y,y_2)$ to drop below 1 at high $p_T$.

\begin{figure}[tbh]
\centering\includegraphics[width=14cm]{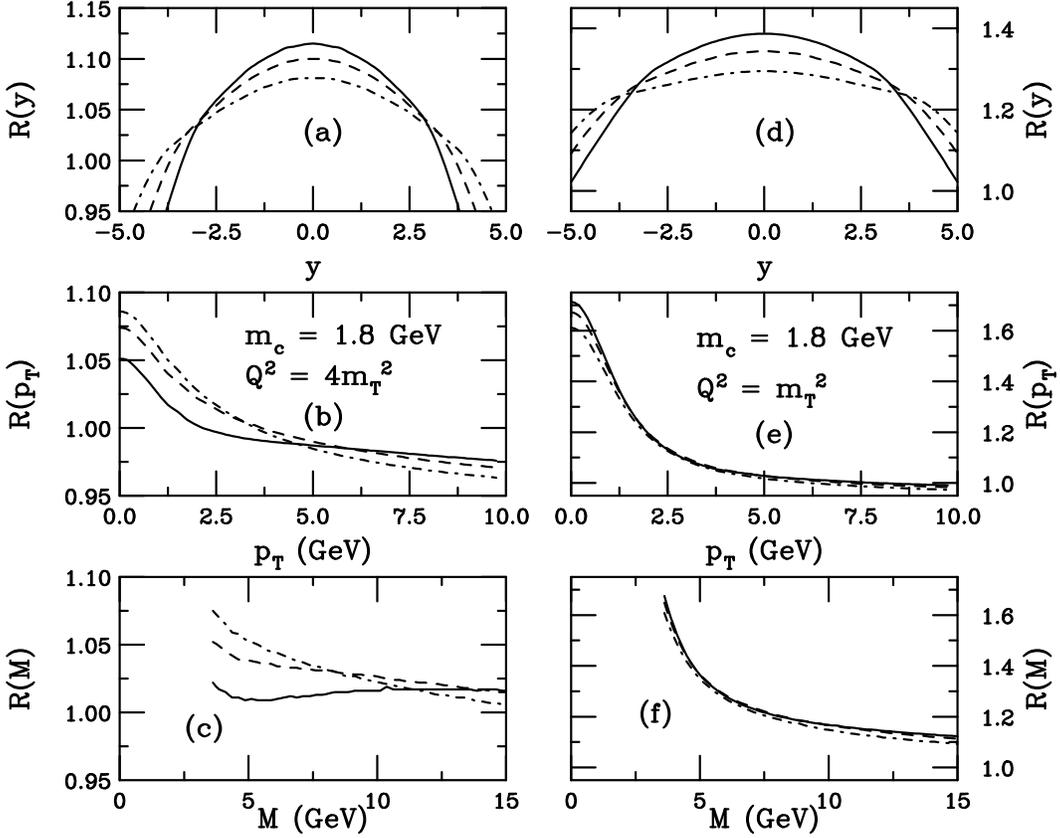}
\caption[]{\small 
As in Fig.~\protect\ref{1213edep} but for $m_c = 1.8$ GeV and $Q^2 = 4m_T^2$
(left-hand side) and $Q^2 = m_T^2$ (right-hand side).
}
\label{18emudep}
\end{figure}

Next, we turn to the integrated ratios,
\begin{equation}
R(y)  \equiv \frac{d\sigma({\scriptstyle\rm EHKQS})/dy}
                  {d\sigma({\scriptstyle\rm CTEQ61L})/dy},\quad
R(p_T)\equiv \frac{d\sigma({\scriptstyle\rm EHKQS})/dp_T}
                  {d\sigma({\scriptstyle\rm CTEQ61L})/dp_T},\quad
R(M)  \equiv \frac{d\sigma({\scriptstyle\rm EHKQS})/dM}
                  {d\sigma({\scriptstyle\rm CTEQ61L})/dM},
\end{equation}
shown in Figs.~\ref{1213edep} and \ref{18emudep}.  All three energies,
$\sqrt{S} = 14$ (solid), 8.8 (dashed) and 5.5 (dot-dashed) TeV, are
shown in each plot. Figure~\ref{1213edep} shows the integrated ratios
for the mass and scale values that best agree with the total cross
section data, $m_c = 1.2$ GeV, $Q^2 = 4m_T^2$ on the left-hand side
and $m_c = 1.3$ GeV, $Q^2 = m_T^2$ on the right-hand side.
Figure~\ref{18emudep} shows the corresponding results for $m_c = 1.8$
GeV with $Q^2 = 4m_T^2$ (left-hand side) and $m_T^2$ (right-hand
side).

As discussed previously, at midrapidity the results for $R(y)$ are
insensitive to the EHKQS extrapolation region $x < x_{\rm min}^{\rm
EHKQS}$.  The magnitude of the enhancement in $R(y)$ can be understood
from the unintegrated results. Since $R(y)$ is integrated over $p_T$,
it does not only reflect the enhancement at $m_T = m_c$ ($p_T=0$)
because $Q^2 \propto m_T^2$ \cite{RVkfac1} and the $p_T$ distribution
peaks around $p_T \approx 1$ GeV. When $R(y)$ is calculated with $Q^2
= m_T^2$, shown on the right-hand sides of the figures, the ratios are
broad due to the stronger growth of $R_g(x_2,Q^2)$ relative to the
reduction of $R_g(x_1,Q^2)$ with increasing $y$, i.e. decreasing $x_2$
and increasing $x_1$.  The ratio is broad because the CTEQ61L gluon
distribution is relatively flat as a function of $x$ for $Q^2 \sim
2-3$ GeV$^2$.  The enhancement decreases and broadens with decreasing
energy.

The rather sharp turnover in $R(y)$ indicates where the extrapolation
region $x<10^{-5}$ begins to contribute.  The rapidity at which
$x<10^{-5}$ is larger for lower energies and larger $m_c$, see
Eq.~(\ref{ylimits}).  The clear decrease of $R(y)$ below 1 at large
rapidity for $Q^2 = 4m_T^2$ is perhaps surprising.  This effect is a
consequence of the small-$x$ extrapolation adopted for the PDFs and
can be understood through an examination of Fig.~\ref{gluons}.  While
the EHKQS gluon distribution is fixed at its value at $x_{\rm
min}^{\rm EHKQS} = 10^{-5}$, the CTEQ61L distribution continues to
change until $x_{\rm min}^{\rm CTEQ61L} = 10^{-6}$.  At $\sqrt{S} =
14$ TeV, the rapidity at which some $x$ values fall below $x_{\rm
min}^{\rm EHKQS}$ is $y_u= 2.2$ (2.6) for $m_c=1.2$ (1.8) GeV. For
$|y|>y_u$ and $Q^2 > 4$ GeV$^2$, as $x$ decreases, the CTEQ61L gluon
distribution increases considerably above the EHKQS distribution.
Thus $R(y) < 1$ at large rapidities.

Since the rapidity distributions are rather flat, there are still
important contributions to the $p_T$ and mass distributions, up to
$\sim30$\% and 40\% respectively at $\sqrt S=14$ TeV for $m_c=1.2$ and
$Q^2=4m_c^2$, from the extrapolation region. The contribution from the
extrapolation region to $R(M)$ is larger since when $m_T^2$ is small,
$M$ can still be large because the difference $y-y_2$ can be large
while either or both $y$ and $y_2$ can be in the low $x$ region.  Thus
the sensitivity of $R(p_T)$ and $R(M)$ to the unconstrained region
should be kept in mind.

The magnitudes of $R(p_T)$ and $R(M)$ in Figs.~\ref{1213edep}
and \ref{18emudep} are similar to those of $R(p_T,y,y_2)$
in Figs.~\ref{1213ydep} and \ref{18ydep}. However, the relative energy
dependence is reversed from the large to small scales.  
At the largest $\sqrt{S}$, the contribution from the $x<10^{-5}$
region is the largest, and as seen in Fig.~\ref{gluons}, if the scales
are {\it e.g.}\ above 5 GeV$^2$, the CTEQ61L gluon density becomes
higher than the EHKQS gluon density at $x \leq 10^{-5}$, suppressing
$R(p_T)$ relative to $R(p_T,y,y_2)$.  At smaller $\sqrt{S}$ the higher
$x$ values reduce the contribution to $d\sigma/dp_T$ from the region
$x<10^{-5}$.  Thus the suppression of $R(p_T)$ from the extrapolation
region is reduced with decreasing energy, explaining the relative
values of $R(p_T)$ and $R(M)$ with energy shown with $Q^2 = 4m_T^2$
(the left-hand sides of Figs.~\ref{1213edep} and \ref{18emudep}).  In
contrast, for $Q^2 = m_T^2$, the EHKQS gluon density is always higher
than CTEQ61L at small $x$ (see Fig. \ref{gluons}), and the energy
dependence remains the same as in Figs.~\ref{1213ydep} and
\ref{18ydep}.

\section{Discussion}

Including the GLRMQ terms with DGLAP evolution enhances the
low $x$ proton gluon distribution \cite{EHKQS}. It may,
however, be very difficult to distinguish between linear and
nonlinear $Q^2$ evolution on the basis of the $F_2$ 
HERA data alone.  Other probes of the low $x$ gluon distribution, 
such as charm production in DIS at HERA and in $pp$ collisions at the
LHC, will hopefully provide the necessary further constraints. In this
paper, we have demonstrated how the nonlinear PDFs {\em
constrained by the HERA data} can cause a significant enhancement in
the LO charm quark cross sections in $pp$ collisions at the LHC.  The
enhancement is defined relative to the results expected with
the pure DGLAP PDFs which also fit the same HERA data. Quantitatively,
however, this enhancement is shown to be very sensitive to the charm
quark mass and the scale.  Clearly, in addition to
collecting more data, further theoretical input, such as high energy
resummation of the heavy quark cross sections \cite{Collins:1991ty}, 
is needed to reduce this sensitivity.

Our basic message is the following.  Currently the high-precision HERA data 
on the structure function 
$F_2(x,Q^2)$ and its derivative with respect to $Q^2$ provide quite stringent
constraints on the gluon distribution. Consequently, the CTEQ61L and 
MRST2001LO gluon distributions are very similar. If a
significant enhancement of charm production, unexpected from DGLAP evolution
alone, is found at the LHC, such an enhancement cannot be absorbed into the 
DGLAP-evolved gluon distributions without introducing a steeper
$\ln Q^2$ slope of $F_2(x,Q^2)$ than allowed by the HERA data.  
Therefore, such an enhancement would be a signal 
of nonlinear effects on the PDF evolution. 

The unintegrated ratios, $R(p_T,y,y_2)$, at $|y,y_2| \leq 2$ and $p_T
\approx 0$ are quite large for the masses and scales that best agree
with the fixed target data. As discussed above, in charm production 
at central rapidities the PDFs are constrained by the HERA
data.  For the smaller charm mass and larger scale, $m_c = 1.2$ GeV
and $Q^2 = 4m_T^2$, we find $R(p_T,y,y_2) \approx 1.4-1.5$.  The
enhancement is even larger for the smaller $Q^2$, $\approx 3.7-5.5$
with $m_c = 1.3$ GeV and $Q^2 = m_T^2$.  The corresponding enhancement
factors are smaller for the upper limit on the charm mass, $m_c = 1.8$
GeV, $1.1-1.2$ for $Q^2=4m_T^2$ and $1.7-2.1$ for $Q^2=m_T^2$.  We
have also shown that the enhancement disappears with increasing $p_T$
due to the decreasing importance of the nonlinear terms at larger
scales in the PDF $Q^2$ evolution.

The enhancements of the integrated ratios are somewhat reduced from
the unintegrated results above. 
The rapidity enhancement is $R(y) \approx 1.07 - 1.8$ at
midrapidity for the energies studied.  For the $p_T$ enhancement, we find
$R(p_T=0) \approx 1.05-1.25$ with $Q^2 = 4m_T^2$ and $\approx
1.6-4.5$ with $Q^2 = m_T^2$ over the range of $m_c$ we investigate.
The mass dependent enhancement, $R(M = 2m_c)$, is similar albeit a bit
smaller than $R(p_T=0)$.  At larger $p_T$ and $M$, the effect dies out
rather quickly, see Figs. \ref{1213edep} and \ref{18emudep}. The main
uncertainty in $R(p_T)$ and $R(M)$ is the sizable contribution from
the region $x<10^{-5}$, currently unconstrained by the HERA data. Thus
the enhancement also depends on the extrapolation of the PDFs in this
$x$ region.

The enhancement we calculate here should be an upper limit on the
possible effect at LO, as we now discuss.  The recombination radius of
the gluon ladders in the GLRMQ terms was assumed to be the proton
radius in the EHKQS analysis \cite{EHKQS}.  The proton radius was the
lower limit on the recombination radius since it gave the strongest
possible recombination effect while still describing the H1 data
\cite{H1} over the full range, $x\ge 3 \times 10^{-5}$ and $Q^2\ge
1.5$~GeV$^2$, without entering the saturation region.  In the
saturation region, further nonlinear terms become important and the
DGLAP+GLRMQ evolution breaks down.  Thus, in this approximation, the
EHKQS gluon distribution is the upper limit on the LO gluon
distribution.  The results obtained here are then an upper limit on
the enhancement.  Studies of the nonlinear PDFs, particularly in the
context of NLO DGLAP evolution, are needed to go beyond this
approximation.

Finally, we discuss possible detection of this
enhancement. Fragmentation and decay should not wipe out the effect,
as also seen for shadowing in $pA$ collisions \cite{ekv1}.  Although
the enhancement $R(p_T,y,y_2)$ is largest, see Figs.~\ref{1213ydep}
and \ref{18ydep}, dileptons may not be the best channel to measure the
enhancement because the origin of the individual pairs is unknown.  The lepton
$p_T$ and $y$ also does not correspond directly to the quark $p_T$ and $y$.  In
addition, it is not clear how the enhancement survives a minimum $p_T$
cut of the charm decay leptons, even though these decay leptons are a
major component of the dilepton spectrum.  Further simulations of this
would be worthwhile.  Since the enhancement disappears at large $p_T$,
the smallest possible $p_T$ cut is desirable.  At higher lepton pair
masses, $b\overline b$ decays dominate the dilepton mass
distributions.  The enhancement rapidly decreases with increasing
quark mass so that effects on $b\overline b$ production are very
small.

Single leptons from charm decays \cite{PHENIX} may be a better
possibility.  The single lepton enhancement may, however, be somewhat
reduced relative to that of the pair.  Of course fully reconstructed
$D$ mesons would be the most desirable option.  Reconstruction should
be possible for $p_T \approx 1$ GeV or less in ALICE with identified
kaons \cite{hpcreport}. More complete simulations with nonlinear PDFs
should thus be performed.

Ideally, a better place to search for the nonlinear effects would be
$pA$ collisions at the LHC \cite{pAHPC}.  The GLRMQ corrections are
expected to get an $\propto A^{1/3}$ enhancement from the nuclear
size, causing the nonlinearities to be significant at larger values of
$x$ and $Q^2$ than in free protons, see Ref.~\cite{Satur} in
Ref.~\cite{pAHPC}. Thus the enhancement should be more pronounced for
nuclei. Before computing charm cross sections in nuclear collisions,
however, the nuclear PDFs should be analyzed within the DGLAP+GLRMQ
framework, including constraints from nuclear deep inelastic
scattering.

\bigskip\bigskip
\noindent {\bf Acknowledgments:}
K.J.E. and V.J.K. gratefully acknowledge the financial support 
from the Academy of Finland, projects 50338 and 80385.
The work of R.V. was supported in part by the Director, Office of
Energy Research, Division of Nuclear Physics of the Office of High
Energy and Nuclear Physics of the U. S.  Department of Energy under
Contract Number DE-AC03-76SF00098.  Support for R.V.'s visit to
Jyv{\"a}skyl{\"a} was provided by the Academy of Finland project 50338.


\begin{thebibliography}{99}

\bibitem{CTEQ6}
%\cite{Pumplin:2002vw}
%\bibitem{Pumplin:2002vw}
J.~Pumplin, D.~R.~Stump, J.~Huston, H.~L.~Lai, P.~Nadolsky and W.~K.~Tung,
%``New generation of parton distributions with uncertainties from global  QCD analysis,''
JHEP {\bf 0207} (2002) 012
[arXiv:hep-ph/0201195].
%%CITATION = HEP-PH 0201195;%%

\bibitem{cteq61}
%\cite{Stump:2003yu}
%\bibitem{Stump:2003yu}
D.~Stump, J.~Huston, J.~Pumplin, W.~K.~Tung, H.~L.~Lai, S.~Kuhlmann and J.~F.~Owens,
%``Inclusive jet production, parton distributions, and the search for new  physics,''
arXiv:hep-ph/0303013.
%%CITATION = HEP-PH 0303013;%%

\bibitem{MRST2001}
%\bibitem{Martin:2001es}
A.~D.~Martin, R.~G.~Roberts, W.~J.~Stirling and R.~S.~Thorne,
%``MRST2001: Partons and alpha(s) from precise deep inelastic scattering  and Tevatron jet data,''
Eur.\ Phys.\ J.\ C {\bf 23} (2002) 73
[arXiv:hep-ph/0110215].
%%CITATION = HEP-PH 0110215;%%

\bibitem{MRSTNNLO}
%\cite{Martin:2002dr}
%\bibitem{Martin:2002dr}
A.~D.~Martin, R.~G.~Roberts, W.~J.~Stirling and R.~S.~Thorne,
%``NNLO global parton analysis,''
Phys.\ Lett.\ B {\bf 531} (2002) 216
[arXiv:hep-ph/0201127].
%%CITATION = HEP-PH 0201127;%%

\bibitem{Martin:2003sk}
A.~D.~Martin, R.~G.~Roberts, W.~J.~Stirling and R.~S.~Thorne,
%``Uncertainties of predictions from parton distributions. II: Theoretical errors,''
arXiv:hep-ph/0308087.
%%CITATION = HEP-PH 0308087;%%

\bibitem{DGLAP} 
%\cite{Dokshitzer:sg}
%\bibitem{Dokshitzer:sg}
Y.~L.~Dokshitzer,
%``Calculation Of The Structure Functions For Deep Inelastic Scattering And E+ E- Annihilation By Perturbation Theory In Quantum Chromodynamics. (In Russian),''
Sov.\ Phys.\ JETP {\bf 46} (1977) 641
[Zh.\ Eksp.\ Teor.\ Fiz.\  {\bf 73} (1977) 1216];
%%CITATION = SPHJA,46,641;%%
%\cite{Gribov:ri}
%\bibitem{Gribov:ri}
V.~N.~Gribov and L.~N.~Lipatov,
%``Deep Inelastic E P Scattering In Perturbation Theory,''
Yad.\ Fiz.\  {\bf 15} (1972) 781
[Sov.\ J.\ Nucl.\ Phys.\  {\bf 15} (1972) 438];
%%CITATION = YAFIA,15,781;%%
%\cite{Gribov:rt}
%\bibitem{Gribov:rt}
V.~N.~Gribov and L.~N.~Lipatov,
%``E+ E- Pair Annihilation And Deep Inelastic E P Scattering In Perturbation Theory,''
Yad.\ Fiz.\  {\bf 15} (1972) 1218
[Sov.\ J.\ Nucl.\ Phys.\  {\bf 15} (1972) 675];
%%CITATION = YAFIA,15,1218;%%
%\cite{Altarelli:1977zs}
%\bibitem{Altarelli:1977zs}
G.~Altarelli and G.~Parisi,
%``Asymptotic Freedom In Parton Language,''
Nucl.\ Phys.\ B {\bf 126} (1977) 298.
%%CITATION = NUPHA,B126,298;%%

%  Yu. Dokshitzer, Sov. Phys. JETP {\bf 46} (1977) 641; 
%  V.N.~Gribov and L.N.~Lipatov, Sov. Nucl. Phys. {\bf 15} (1972) 438, 675; 
%  G.~Altarelli, G.~Parisi, Nucl. Phys. {\bf B126} (1977) 298.

\bibitem{GLR}
%\cite{Gribov:ac}
%\bibitem{Gribov:ac}
L.~V.~Gribov, E.~M.~Levin and M.~G.~Ryskin,
%``Singlet Structure Function At Small X: Unitarization Of Gluon Ladders,''
Nucl.\ Phys.\ B {\bf 188} (1981) 555;
%%CITATION = NUPHA,B188,555;%%
%\cite{Gribov:tu}
%\bibitem{Gribov:tu}
L.~V.~Gribov, E.~M.~Levin and M.~G.~Ryskin,
%``Semihard Processes In QCD,''
Phys.\ Rept.\  {\bf 100} (1983) 1.
%%CITATION = PRPLC,100,1;%%

\bibitem{MQ}
%\cite{Mueller:wy}
%\bibitem{Mueller:wy}
A.~H.~Mueller and J.~w.~Qiu,
%``Gluon Recombination And Shadowing At Small Values Of X,''
Nucl.\ Phys.\ B {\bf 268} (1986) 427.
%%CITATION = NUPHA,B268,427;%%

\bibitem{EHKQS}
%\bibitem{Eskola:2002yc}
K.~J.~Eskola, H.~Honkanen, V.~J.~Kolhinen, J.~w.~Qiu and C.~A.~Salgado,
%``Nonlinear corrections to the DGLAP equations in view of the HERA data,''
Nucl.\ Phys.\ B {\bf 660} (2003) 211
[arXiv:hep-ph/0211239].
%%CITATION = HEP-PH 0211239;%%

\bibitem{H1}
%\bibitem{Adloff:2000qk}
C.~Adloff {\it et al.}  [H1 Collaboration],
%``Deep-inelastic inclusive e p scattering at low x and a determination of  alpha(s),''
Eur.\ Phys.\ J.\ C {\bf 21} (2001) 33
[arXiv:hep-ex/0012053].
%%CITATION = HEP-EX 0012053;%%

\bibitem{RVkfac}
%\cite{Vogt:2002eu}
%\bibitem{Vogt:2002eu}
R.~Vogt,
%``What is the real K factor?,''
Heavy Ion Phys.\  {\bf 17} (2003) 75
[arXiv:hep-ph/0207359].
%%CITATION = HEP-PH 0207359;%%

\bibitem{COMBRIDGE}
%\cite{Combridge:1978kx}
%\bibitem{Combridge:1978kx}
B.~L.~Combridge,
%``Associated Production Of Heavy Flavor States In P P And Anti-P P Interactions: Some QCD Estimates,''
Nucl.\ Phys.\ B {\bf 151} (1979) 429.
%%CITATION = NUPHA,B151,429;%%

%\bibitem{Lai:1999wy}
\bibitem{CTEQ5}
H.~L.~Lai {\it et al.}  [CTEQ Collaboration],
%``Global {QCD} analysis of parton structure of the nucleon: CTEQ5 parton  distributions,''
Eur.\ Phys.\ J.\ C {\bf 12} (2000) 375
[arXiv:hep-ph/9903282].
%%CITATION = HEP-PH 9903282;%%

\bibitem{cteqwww}
  http://user.pa.msu.edu/wkt/cteq/cteq6/cteq6pdf.html

\bibitem{mrst}
%\cite{Martin:1998sq}
%\bibitem{Martin:1998sq}
A.~D.~Martin, R.~G.~Roberts, W.~J.~Stirling and R.~S.~Thorne,
%``Parton distributions: A new global analysis,''
Eur.\ Phys.\ J.\ C {\bf 4} (1998) 463
[arXiv:hep-ph/9803445];
%%CITATION = HEP-PH 9803445;%%
%\cite{Martin:1998np}
%\bibitem{Martin:1998np}
A.~D.~Martin, R.~G.~Roberts, W.~J.~Stirling and R.~S.~Thorne,
%``Scheme dependence, leading order and higher twist studies of MRST  partons,''
Phys.\ Lett.\ B {\bf 443} (1998) 301
[arXiv:hep-ph/9808371].
%%CITATION = HEP-PH 9808371;%%

\bibitem{HPC}
%\cite{Vogt:2001nh}
%\bibitem{Vogt:2001nh}
R.~Vogt  [Hard Probe Collaboration],
%``The A dependence of open charm and bottom production,''
Int.\ J.\ Mod.\ Phys.\ E {\bf 12} (2003) 211
[arXiv:hep-ph/0111271].
%%CITATION = HEP-PH 0111271;%%

\bibitem{rvww02}
 R. Vogt, in proceedings of the {\it 18$^{\rm th}$ Winter Workshop on Nuclear
Dynamics}, edited by R. Bellwied {\it et al.}, Nassau, The Bahamas, 2002, 
p. 253.

\bibitem{NDE}
P. Nason, S. Dawson, and R. K. Ellis, Nucl. Phys. {\bf B327} (1989) 49.

\bibitem{RVkfac1}
%\cite{Vogt:1995zf}
%\bibitem{Vogt:1995zf}
R.~Vogt,
%``Phenomenology of Charm and Bottom Production,''
Z.\ Phys.\ C {\bf 71} (1996) 475
[arXiv:hep-ph/9510293].
%%CITATION = HEP-PH 9510293;%%

%\cite{Collins:1991ty}
\bibitem{Collins:1991ty}
J.~C.~Collins and R.~K.~Ellis,
%``Heavy quark production in very high-energy hadron collisions,''
Nucl.\ Phys.\ B {\bf 360} (1991) 3.
%%CITATION = NUPHA,B360,3;%%

\bibitem{ekv1}
%\bibitem{Eskola:2001gt}
K.~J.~Eskola, V.~J.~Kolhinen and R.~Vogt,
%``Obtaining the nuclear gluon distribution from heavy quark decays to  lepton pairs in p A collisions,''
Nucl.\ Phys.\ A {\bf 696} (2001) 729
[arXiv:hep-ph/0104124].
%%CITATION = HEP-PH 0104124;%%

\bibitem{PHENIX}
%\cite{Adcox:2002cg}
%\bibitem{Adcox:2002cg}
K.~Adcox {\it et al.}  [PHENIX Collaboration],
%``Measurement of single electrons and implications for charm production  in Au + Au collisions at s(NN)**(1/2) = 130-GeV,''
Phys.\ Rev.\ Lett.\  {\bf 88} (2002) 192303
[arXiv:nucl-ex/0202002].
%%CITATION = NUCL-EX 0202002;%%


\bibitem{hpcreport}
    ALICE Collaboration, Technical Proposal, CERN/LHCC 95-71;
    ALICE Collaboration, Addendum to the Letter of Intent, CERN/LHCC 95-24;
    ALICE Collaboration, Addendum to ALICE Proposal, CERN/LHCC 99-13;
    ALICE Collaboration, Hard Probes in Heavy Ion Collisions at the LHC.

\bibitem{pAHPC}
%\bibitem{Accardi:2003be}
A.~Accardi {\it et al.},
``Hard probes in heavy ion collisions at the LHC: PDFs, shadowing and $pA$ collisions,'' ed. K.J. Eskola, arXiv:hep-ph/0308248.
%%CITATION = HEP-PH 0308248;%%

%\cite{Eskola:2003gc}
%\bibitem{Eskola:2003gc}
\bibitem{Satur}
K.~J.~Eskola, H.~Honkanen, V.~J.~Kolhinen, J.~w.~Qiu and C.~A.~Salgado,
%``Nonlinear corrections to the DGLAP equations: Looking for the  saturation limits,''
arXiv:hep-ph/0302185.
%%CITATION = HEP-PH 0302185;%%



\end{thebibliography}
\end{document}